\documentclass{article}
\begin{document}
\title{Irreversibility in the Halting Problem  of Quantum
Computer}
\author{A.E.Shalyt-Margolin, V.I.Strazhev and A.Ya.Tregubovich \\
National Centre of High Energy and Particle Physics,\\
 Bogdanovich Str. 153, 220040 Minsk, Belarus,\\
E-mails: alexm@hep.by, a.tregub@open.by}
\date{}
\maketitle

\begin{abstract}
  The Halting problem of a quantum computer is considered. It is
shown that if halting  of a quantum computer takes place the
associated dynamics is described by an irreversible operator.
\end{abstract}

\section{Introduction}
In the last 10 - 15 years a quantum theory of information
has developed rather intensively in various directions from
mathematical aspects to different physical problems: quantum
algorithms \cite{r1} , quantum decoherence \cite{r2}, density
matrix and entropy for the entanglement states \cite{r3}, measuring
theory for quantum information, and a number of physical models
of quantum computers based on various principles \cite{r4}.

However, there are problems to be solved. Among them is
the halting problem that has originated in the mid eighties. This problem may
be generally formulated as follows: how can a correct description of the
quantum computer halting be compatible with the basic principles of
a quantum theory of information \cite{r5}.

This problem is studied in a number of works \cite{r6}--\cite{r9},
\cite{r12}, \cite{r13}. And in the present paper it is shown that halting
of quantum computers is incompatible not only with unitarity but  also with
reversibility of the corresponding dynamics.

\section{The Halting problem}
  In the paper \cite{r5}, where the term 'halting' is
firstly used, the following special qubit is chosen
 $$
\hat{q} = \left(
\begin{array}{c}
0 \\
1
\end{array}
\right)
 $$
to signal that the computer is halted. This means that each correctly
working program sets
$\hat{q}$ to $1$ when then operation is terminated, and sets
$\hat{q}$ to $0$ otherwise. According to Myers \cite{r6}, the program
for different branches of the computing
process can have different number of  steps giving rise to the unitarity problem
of the basic calculation operators.
This problem was in principal solved in \cite{r7}.
But as shown by \cite{r8}, in \cite{r7} a kind of the Turing machine is used,
inapplicable to realistic computers, as in this case the
dynamics is unitary only for the computers having no  halt. So
for realistic computers the problem remains unsolved. Besides,
in \cite{r8} one more problem is discovered for
a quantum computer when different branches of the computation
process halt at different and unknown times. And in
\cite{r9} it is shown that halting of
the universal quantum computer is incompatible with the unitarity constraint
of quantum computations.

\section{The Halting problem, unitarity and reversibility}

To make the following definitions valid, we use the terminology of \cite{r9}:
\begin{enumerate}
\item Quantum computer is a closed quantum system controlled by the time -
independent evolution operator $U$ for each time step between the state of the
input space representing some vector $\mid \tau _{in}\rangle $ in a Hilbert space  $
\hat{H}$ and the final state $\mid \tau _{out}\rangle $ of the output in the same
Hilbert space.
\item For halting, the dynamics is to be able to store
the output that is finite in terms of qubit resources, no matter
in what finite time  the desirable output is
computed. This reserved space, from where the output can be
read out, is mathematically an invariant
subspace $V \subset \hat{H}$ with a component of the qubit
$\hat{q}$ equal to $1$.
\end{enumerate}
We intentionally weaken the requirements for the dynamics and do
not consider $U$ as obviously unitary.

Thus, any state has the form
\begin{equation}\label{1}
\left| \psi_{0}\right.\rangle = \left| 0_{h}\right.\rangle \otimes
\left| x_{0}\right.\rangle + \left| 1_{h}\right.\rangle \otimes
\left| y_{0}\right.\rangle
\end{equation}
The information transfer matrix $U$ is written as
\begin{equation}\label{2}
U = \left(
\begin{array}{cc}
A & \alpha \\
0 & B
\end{array}
\right)
\end{equation}
in the basis
\begin{equation}\label{3}
\mid 0_{h}\rangle \otimes \mid x_{0}\rangle = \left(
\begin{array}{c}
0 \\
\mid x_{0}\rangle
\end{array}
\right) , \mid 1_{h}\rangle \otimes \mid y_{0}\rangle =
\left(
\begin{array}{c}
\mid y_{0}\rangle \\
0
\end{array}
\right)
\end{equation}
We show that the halting conditions of a quantum computer after
performance of the program with a finite number of steps
\begin{equation}\label{41}
 {\rm for}\;  N \geq N_{0} \quad U^{N}\mid \psi
_{0} \rangle = \mid 1_{h}\rangle \otimes \mid y_{0}\rangle
\end{equation}
\begin{equation}\label{42}
\left\langle 0_{h}\mid U^{N}\mid \psi _{0}\right\rangle = B^{N}\mid x_{0}
\rangle = 0
\end{equation}
are incompatible with the reversibility of the operator $U$.

Actually, let $U$ be a two-side reversible matrix and let
$$
^{-1}U =
\left(
\begin{array}{cc}
A_{11}^{(l)} & A_{12}^{(l)} \\
A_{21}^{(l)} & A_{22}^{(l)}
\end{array}
\right)
$$
 be the left-hand reciprocal for $U$, whereas
 $$
 U^{-1} =\left(
\begin{array}{cc}
A_{11}^{(r)} & A_{12}^{(r)} \\
A_{21}^{(r)} & A_{22}^{(r)}
\end{array}
\right)
$$
 be the right-hand reciprocal for $U$.  Then
\begin{equation}\label{5}
^{-1}UU = \left(
\begin{array}{cc}
A_{11}^{(l)} & A_{12}^{(l)} \\
A_{21}^{(l)} & A_{22}^{(l)}
\end{array}
\right) \left(
\begin{array}{cc}
A & \alpha \\
0 & B
\end{array}
\right) =\left(
\begin{array}{cc}
A_{11}^{(l)}A & A_{11}^{(l)}\alpha +A_{12}^{(l)}B \\
A_{21}^{(l)}A & A_{21}^{(l)}\alpha +A_{22}^{(l)}B
\end{array}
\right) = \left(
\begin{array}{cc}
1 & 0 \\
0 & 1
\end{array}
\right)
\end{equation}
Thus it follows that the matrix
$A$ has the left-hand reciprocal  $A^{-1}
= A_{11}^{(l)}$.
Similarly, we have
\begin{equation}\label{6}
UU^{-1} = \left(
\begin{array}{cc}
A & \alpha \\
0 & B
\end{array}
\right) \left(
\begin{array}{cc}
A_{11}^{(r)} & A_{12}^{(r)} \\
A_{21}^{(r)} & A_{22}^{(r)}
\end{array}
\right) = \left(
\begin{array}{cc}
AA_{11}^{(r)} + \alpha A_{21}^{(r)} & AA_{12}^{(r)} + \alpha A_{22}^{(r)} \\
BA_{21}^{(r)} & BA_{22}^{(r)}
\end{array}
\right) = \left(
\begin{array}{cc}
1 & 0 \\
0 & 1
\end{array}
\right)
\end{equation}
Consequently, the matrix $B$ is right-hand reversible.
As $B$ is right-hand reversible, $B^{N}$ is such as well.

Using the results for examples 8 and 10 of Chapter 2 from \cite{r10}, we
obtain immediately that $(B^{N})^{+}$ is also right-reversible. The condition
(\ref{42}) is obviously equivalent to the condition
\begin{equation}\label{42a}
\langle x_{0}\mid (B^{N})^{+} = 0.
\end{equation}

Multiplying the left and right parts of the last equality on the right by
$(B^{N}{}^{+})^{-1}$, we obtain for any bra - vector
 $\langle x_{0}\mid = 0$. This is an obvious contradiction.
So it follows that $U$ is not reversible.
Besides from our proof it follows that
$U$ is not even right-hand reversible.
It is necessary to make two remarks:
\begin{enumerate}
\item The above proof is correct both for the universal quantum computer, that is
the case when the Hilbert space $\hat{H}$ is infinitely dimensional, and
for the realistic quantum computer when $\hat{H}$ has a finite
dimension. The proof is simplified in this case due to the fact that (a) for
square matrices a left-hand reciprocal is coincident with the right-hand one and (b)
for the upper triangular reversible matrices the  Jordan decomposition takes
place  \cite{r11}.
$$
 U = \left(
\begin{array}{cc}
A & \alpha \\
0 & B
\end{array}
\right) = \left(
\begin{array}{cc}
A & 0  \\
0 & B
\end{array}
\right) \left(
\begin{array}{cc}
1 & A^{-1}\alpha \\
0 & 1
\end{array}
\right)
$$
Then the key argument will follow directly from the condition of (\ref{42}).
\item It would be natural to require that the left-hand
reciprocal $^{-1}U$ and the right-hand reciprocal $U^{-1}$ of $U$ be also
elements of the dynamics of a quantum computer
and should be of the upper triangular form
to simplify the proof even greater.
\end{enumerate}

\section{Conclusion}

In the work it is shown that in the general case when
halting of the universal or realistic quantum computer takes place the
associated dynamics is non-unitary and, what is more, irreversible.

\section{Acknowlegement}

The author would like to thank the Referee for very helpful comments.


\end{document}